\documentclass[]{agujournal2019}

\usepackage{url} 
\usepackage{amsmath}

\usepackage[normalem]{ulem} 

\journalname{Journal of Advances in Modeling Earth Systems (JAMES)}

\begin{document}

%
%

\title{Machine Learning for Stochastic Parameterization: \\ Generative Adversarial Networks in the Lorenz '96 Model}
%
%
% ...
%

%
%



\newcommand{\note}[1]{\textsf{\color{red} Note: [#1]}}

\authors{David John Gagne II\affil{1}\thanks{PO Box 3000, Boulder, CO 80307},
         Hannah M. Christensen\affil{1,2}, Aneesh C. Subramanian \affil{3},
         Adam H. Monahan \affil{4}}


\affiliation{1}{National Center for Atmospheric Research, Boulder, Colorado, USA}
\affiliation{2}{Atmospheric, Oceanic and Planetary Physics, University of Oxford, UK}
\affiliation{3}{University of Colorado Boulder, Boulder, Colorado, USA}
\affiliation{4}{School of Earth and Ocean Sciences, University of Victoria, Victoria, British Columbia, Canada}




\correspondingauthor{David John Gagne II}{dgagne@ucar.edu}




\begin{keypoints}
\item We have developed a generative adversarial network machine learning stochastic parameterization of sub-grid forcing for the Lorenz '96 dynamical model.
\item We test different configurations of the generative adversarial network parameterization at weather and climate timescales and find that some have representation errors better than a bespoke polynomial parameterization.
\item The generative adversarial networks closely reproduce the spatio-temporal correlations and regimes of the Lorenz '96 system. 
\end{keypoints}

%
%


\begin{abstract}
Stochastic parameterizations account for uncertainty in the representation of unresolved sub-grid processes by sampling from the distribution of possible sub-grid forcings. Some existing stochastic parameterizations utilize data-driven approaches to characterize uncertainty, but these approaches require significant structural assumptions that can limit their scalability. Machine learning models, including neural networks, are able to represent a wide range of distributions and build optimized mappings between a large number of inputs and sub-grid forcings. Recent research on machine learning parameterizations has focused only on deterministic parameterizations. In this study, we develop a stochastic parameterization using the generative adversarial network (GAN) machine learning framework. The GAN stochastic parameterization is trained and evaluated on output from the Lorenz '96 model, which is a common baseline model for evaluating both parameterization and data assimilation techniques. We evaluate different ways of characterizing the input noise for the model and perform model runs with the GAN parameterization at weather and climate timescales. Some of the GAN configurations perform better than a baseline bespoke parameterization at both timescales, and the networks closely reproduce the spatio-temporal correlations and regimes of the Lorenz '96 system. We also find that in general those models which produce skillful forecasts are also associated with the best climate simulations. 
\end{abstract}

\section*{Plain Language Summary}
Simulations of the atmosphere must approximate the effects of small-scale processes with simplified functions called parameterizations. Standard parameterizations only predict one output for a given input, but stochastic parameterizations can sample from all the possible outcomes that can occur under certain conditions. We have developed and evaluated a machine learning stochastic parameterization, which builds a mapping between large-scale current conditions and the range of small-scale outcomes from data about both. We test the machine learning stochastic parameterization in a simplified mathematical simulation that produces multi-scale chaotic waves like the atmosphere. We find that some configurations of the machine learning stochastic parameterization perform slightly better than a simpler baseline stochastic parameterization over both weather- and climate-like time spans.  
%
%

%


%
%
%
%

\section{Introduction}

A large source of weather and climate model uncertainty is the approximate representation of unresolved sub-grid processes through parameterization schemes. Traditional, deterministic parameterization schemes represent the mean or  most likely sub-grid scale forcing for a given resolved-scale state. While model errors can be reduced to a certain degree through improvements to such parameterizations, they cannot be eliminated. Irreducible uncertainties result from a lack of scale separation between resolved and unresolved processes. Uncertainty in weather forecasts also arises because the chaotic nature of the atmosphere gives rise to sensitivity to uncertain initial conditions. Practically, uncertainty is represented in forecasts using ensembles of integrations of comprehensive weather and climate prediction models, first suggested by \citeA{Leith:1975jc}. To produce reliable probabilistic forecasts, the generation of the ensemble must include a representation of both model and initial condition uncertainty.

Initial condition uncertainty is addressed by perturbing the initial conditions of ensemble members, for example by selecting directions of optimal perturbation growth using singular vectors \cite{buizza1995}, or by characterizing initial condition uncertainty during the data assimilation cycle \cite{isaksen2010}. One approach for representing irreducible model uncertainty is stochastic parameterization of unresolved physical processes. A stochastic parameterization represents the probability distribution of possible sub-grid scale tendencies conditioned on the large scale. Each ensemble member experiences one possible, equally likely realization of the sub-grid-scale tendencies. A more detailed motivation for including stochastic parameterizations in weather and climate models is presented in \citeA{Palmer:2012fe}.

Stochastic approaches for numerical weather prediction (NWP) were originally proposed for use in the European Center for Medium-Range Weather Forecasts (ECMWF) ensemble prediction system \cite{Palmer:1997va, Buizza1999}. They were shown to substantially improve the quality of initialized ensemble forecasts, and so became widely adopted by meteorological services around the world, where they are used to produce operational ensemble weather, sub-seasonal and seasonal forecasts \cite{Teixeira:2010hd, Reyes:2009iy, Berner:2010fn, Stockdale:2011km,Palmer:2009ji,Palmer:2012fe, Suselj:2013jd, Suselj:2014th, Weisheimer:2014df, Sanchez2016, leutbecher2017}. 

Recent work has assessed the impact of stochastic parameterization schemes in both idealized and state-of-the-art climate models for long term integration \cite{Williams2012, Ajayamohan2013,Juricke2014,Dawson2015,wang2016,christensen2017,davini2017,strommen2018}. These studies demonstrate that including a stochastic representation of model uncertainty can go beyond improving initialized forecast reliability, and can also lead to improvements in the model mean state \cite{Palmer:2001gw,Berner:2012gx}, climate variability \cite{Ajayamohan2013,Dawson2015,christensen2017}, and change a model's climate sensitivity \cite{Seiffert2010}. These impacts occur through non-linear rectification, noise enhanced variability, and noise-induced regime transitions \cite{berner2017}. In this way, small-scale variability represented by stochastic physics can impact large spatio-temporal scales of climate variability.
   
Despite the historical disconnect between the weather and climate prediction communities, the boundaries between weather and climate prediction are somewhat artificial \cite{Palmer:2008ck,Hurrell2009,Shapiro2010}. This disconnect is highlighted by recent advances in prediction on timescales from weather to sub-seasonal-to-seasonal and decadal by operational weather forecasting centers around the world \cite{Moncrieff2007,Vitart:2012ip}. Nonlinearities in the climate system lead to an upscale transfer of energy (and therefore error) from smaller to larger scales \cite{lorenz1969,Palmer:2001gw,tribbia2004}. At the same time, slowly evolving modes of variability can produce predictable signals on shorter timescales \cite{hoskins2013,vannitsem2016}. Under the `seamless prediction' paradigm, the weather and climate communities should work together to develop Earth-system models \cite{brunet2010,christensen2019}, as developments made in one community are expected to benefit the other. The development and use of stochastic parameterizations is a good example of this paradigm at work.

Recent years have seen substantial interest in the development of stochastic parameterization schemes. Pragmatic approaches, such as the Stochastically Perturbed Parameterization Tendencies (SPPT) scheme \cite{Buizza1999,Palmer:2009uu} are widely used due to their ease of implementation and beneficial impacts on the model \cite{Sanchez2016,leutbecher2017,christensen2017}. Other schemes predict the statistics of model uncertainty using a theoretical understanding of the atmospheric processes involved, such as the statistics of convection \cite{craig2006,khouider2010,sakradzija2018,bengtsson2019}. A third approach is to make use of observations or high-resolution simulations to characterize variability that is unresolved in a low-resolution forecast model \cite{shutts2007}. This last approach can be used to develop data-driven stochastic schemes \cite{dorrestijn2015,bessac2019} or to constrain tunable parameters in stochastic parameterizations \cite{shutts2014,christensen2015d,christensen2019b}. A drawback of these data-driven approaches is that assumptions are made about the structure of the stochastic parameterization (e.g. the physical process to focus on, or the distribution of the stochastic term conditioned on the resolved state) in order to make the analysis tractable using conventional methods.

Machine learning models offer an approach to parameterize complex nonlinear sub-grid processes in a potentially computationally efficient manner from data describing those processes. The family of machine learning models consist of mathematical models whose structure and parameters (often denoted weights) optimize the predictive performance of \emph{a priori} unknown relationships between input (``predictor'') and output (``predictand'') variables. Commonly used machine learning model frameworks range in complexity from simple linear regression to decision trees and neural networks. More complex methods allow modelling of broader classes of predictor-predictand relationships. Machine learning models minimize overfitting through the use of regularization techniques that impose constraints on the model structure and weights. Regularization is critical for more complex machine learning models, so that they can converge to optimal and robust configurations in large parameter spaces. For the parameterization problem, regularization and physical constraints are critical for ensuring that the machine learning model predictions match the distribution of observed values. Machine learning for parameterizations has been considered since \cite{Krasnopolsky2005} and recently multiple groups have begun developing new parameterizations for a variety of processes \cite{Schneider2017, Gentine2018, Rasp2018, Bolton2019}. However, these schemes have focused exclusively on deterministic parameterization approaches, but the need for stochastic perturbations is being recognized \cite{Brenowitz2019}. 

 One active area in current machine learning research is generative modeling, which focuses on models that create synthetic representative samples from a distribution of arbitrary complexity without the need for a parametric representation of the  distribution. Generative adversarial networks, or GANs \cite{Goodfellow2014}, are a class of generative models that consist of two neural networks in mutual competition. The generator network produces synthetic samples similar to the original training data from a latent vector, and the critic, or discriminator, network examines samples from the generator and the training data in order to determine if a sample is real or synthetic. The critic acts as an adaptable loss function for the generator by learning features of the training data and teaching those features to the generator through back-propagation. The original GAN formulation used a latent vector of random numbers as the only input to the generator, but subsequent work on conditional GANs \cite{Mirza2014} introduced the ability to incorporate a combination of fixed and random inputs into the generator, enabling sampling from conditional distributions. Because the stochastic parameterization problem can be framed as sampling from the distribution of sub-grid tendencies conditioned on the resolved state, conditional GANs have the potential to perform well on this task.

The purpose of this study is to evaluate how well GANs can parameterize the sub-grid tendency component of an atmospheric model at weather and climate timescales. A key question is whether a GAN can learn uncertainty quantification within the parameterization framework, removing the need to retrospectively develop separate stochastic representations of model uncertainty. While the ultimate goal is to test these ideas in a full general circulation model (GCM: left for a future work), as a proof of concept we will use the two-timescale model proposed in \citeA{lorenz1996}, henceforth the L96 system, as a testbed for assessing the use of GAN in atmospheric models. Simple chaotic dynamical systems such as L96 are useful for testing methods in atmospheric modeling due to their transparency and computational cheapness. The L96 system has been widely used as a testbed in studies including development of stochastic parameterization schemes \cite{wilks2005,crommelin2008,kwasniok2012,Arnold2013}, data assimilation methodology \cite{fertig2007,law2016,hatfield2018}, as well as using ML approaches to learn improved deterministic parameterization schemes \cite{Schneider2017,dueben2018,watson2019}.

The evaluation consists of four primary questions. First, given inputs from the ``true" L96 model run, how closely does the GAN approximate the true distribution of sub-grid tendency? Second, when an ensemble of L96 models with stochastic GAN parameterizations are integrated forward to medium range weather prediction time scales, how quickly does the prediction error increase and how well does the ensemble spread capture the error growth? Third, when the L96 model with a stochastic GAN parameterization is integrated out to climate time scales, how well does the GAN simulation approximate the properties of the true climate? Fourth, how closely does the GAN represent both different regimes within the system and the probability of switching between them?

Details of the Lorenz '96 model and the GAN are presented in Section 2, and the results of the weather and climate analyses described above are presented in Section 3.  Section 4 provides a discussion of GAN parameterization `health risks', while an overall discussion of results is presented in Section 5.  Conclusions follow in Section 6.

\section{Methods}

\subsection{Lorenz '96 Model Configuration}
 
 The L96 system was designed as a `toy model' of the extratropical atmosphere, with simplified representations of advective nonlinearities and multi-scale interactions  \cite{lorenz1996}. It consists of two scales of variables arranged around a latitude circle. The large scale, low-frequency $X$ variables are coupled to a larger number of small-scale high-frequency $Y$ variables, with a two-way interaction between the $X$s and $Y$s. It is the interaction between variables of different scales that makes the L96 system ideal for evaluating new ideas in parameterization development.  The L96 system has proven useful in assessing new techniques that were later developed for use in GCMs \cite{crommelin2008,dorrestijn2013}.

The $X$ and $Y$ variables evolve following:
\begin{subequations}
\begin{align}
\frac{\mathrm{d}X_k}{\mathrm{d}t} &= -X_{k-1}(X_{k-2} - X_{k+1}) - X_k + F - \frac{hc}{b} \sum_{j = J(k-1)+1}^{kJ} Y_j; \; &k =1, ..., K \\
\frac{\mathrm{d}Y_j}{\mathrm{d}t} &= - c b Y_{j+1} (Y_{j+2} - Y_{j-1}) - cY_j + \frac{hc}{b}X_{(\mathrm{int}[(j-1)/J]+1)}; \; &j = 1, ..., JK,
\end{align}
\label{eq:L96}
\end{subequations}
where in the present study the number of $X$ variables, $K=8$ and the number of $Y$ variables per $X$ variable, $J=32$. Further, we set the coupling constant, $h=1$, the spatial-scale ratio, $b=10$ and the temporal-scale ratio $c=10$. 
The forcing term $F=20$ is set large enough to ensure chaotic behavior. The chosen parameter settings, which were used in \cite{Arnold2013}, are such that one model time unit (MTU) is approximately equivalent to five atmospheric days, deduced by comparing error-doubling times in L96 and the atmosphere \cite{lorenz1996}.

In this study the full Lorenz `96 equations are treated as the `truth' which must be forecast or simulated. In the case of the atmosphere, the physical equations of motion of the system are known. However, due to limited computational resources, it is not possible to explicitly simulate the smallest scales, which are instead parameterized. Motivated by this requirement for weather and climate prediction, a forecast model for the L96 system is constructed by truncating the model equations, and parameterizing the impact of the small $Y$ scales on the resolved $X$ scales:
\begin{equation}
\frac{\mathrm{d}X^*_k}{\mathrm{d}t} = -X^*_{k-1}(X^*_{k-2} - X^*_{k+1}) - X^*_k + F - \hat{U}(X^*_k,t); \; k =1, ..., K, 
\label{eq:F96}
\end{equation}
where $X^*_k(t)$ is the forecast estimate of $X_{k}(t)$ and $\hat{U}(X^*_k,t)$ is the parameterized sub-grid tendency. The parameterization $\hat{U}$ approximates the true sub-grid tendencies:
\begin{equation}
U(X,Y) = \frac{hc}{b} \sum_{j = J(k-1)+1}^{kJ} Y_j,
\end{equation}
which can be estimated from realizations of the ``truth'' time series as
\begin{equation}
U_k(t) = [-X_{k-1}(t)\left(X_{k-2}(t) - X_{k+1}(t)\right) - X_{k}(t) + F] - \left(\frac{X_{k}(t+dt_f) - X_{k}(t)}{dt_f}\right).
\label{eqn:truthU}
\end{equation}
following \citeA{Arnold2013}. The time step $dt_f$ equals the time step used in the forecast model for consistency.

A long ``truth'' run of the L96 model is performed to generate both training data for the machine learning models and a test period for both weather and climate evaluations. The ``truth'' run is integrated for 20000 MTU using a fourth-order Runge-Kutta timestepping scheme and a time step $dt=0.001$ MTU. Output from the first 2000 MTU are used for training, and the remaining 18000 MTU are used for testing. A burn-in period of 2 MTU is discarded. All parameterized forecast models of the L96 use a forecast timestep of $dt_f=0.005$ MTU and a second order Runge-Kutta timestepping scheme. This scheme is to represent the time discretization of the equations representing the resolved dynamics in an atmospheric forecasting model.

\subsection{GAN Parameterizations}
 
The GAN parameterization developed for the Lorenz '96 model in this study utilizes a conditional dense GAN to predict the sub-grid tendency at the current time step given information about the state at the previous time step. We will investigate two classes of predictors of $U_{t}$: both $X$ and $U$ at the previous forecast timestep, and $X$ alone. In the following discussion we focus on GANs based on the first of these predictor sets; the construction of those associated with the second set is analogous to that of the first. Note that in this section we move to a discrete-time notation, with the time index indicated by the subscript, $t$, where adjacent indices are separated by the forecast time step, $dt_f$.

The GAN generator accepts $X_{t-1, k}$, $U_{t-1, k}$, and a latent random vector $Z_{t-1, k}$ as input to estimate $\hat{U}_{t, k}$, or the predicted $U$ at time $t$. The discriminator accepts $X_{t-1, k}$, $U_{t-1, k}$, and $V_{t, k}$ as inputs (where $V_{t,k}$ may be either $U_{t,k}$ if from the training data or $\hat{U}_{t,k}$ if from the generator) and outputs the probability that $V_{t, k}$ comes from the training data. All inputs and outputs are re-scaled to have a mean of 0 and standard deviation of 1 based on the training data distributions. Note that we choose to develop local GANs, i.e. GANs which accept $X$ and $U$ for a given spatial index, $k$, and that predict $\hat{U}$ for that index, $k$, as opposed to GANs that accept vector $X$ and $U$ and thus include spatial information. This is to match the local nature of parameterization schemes in weather and climate models. 

\begin{figure}
\centering
\includegraphics[width=\textwidth]{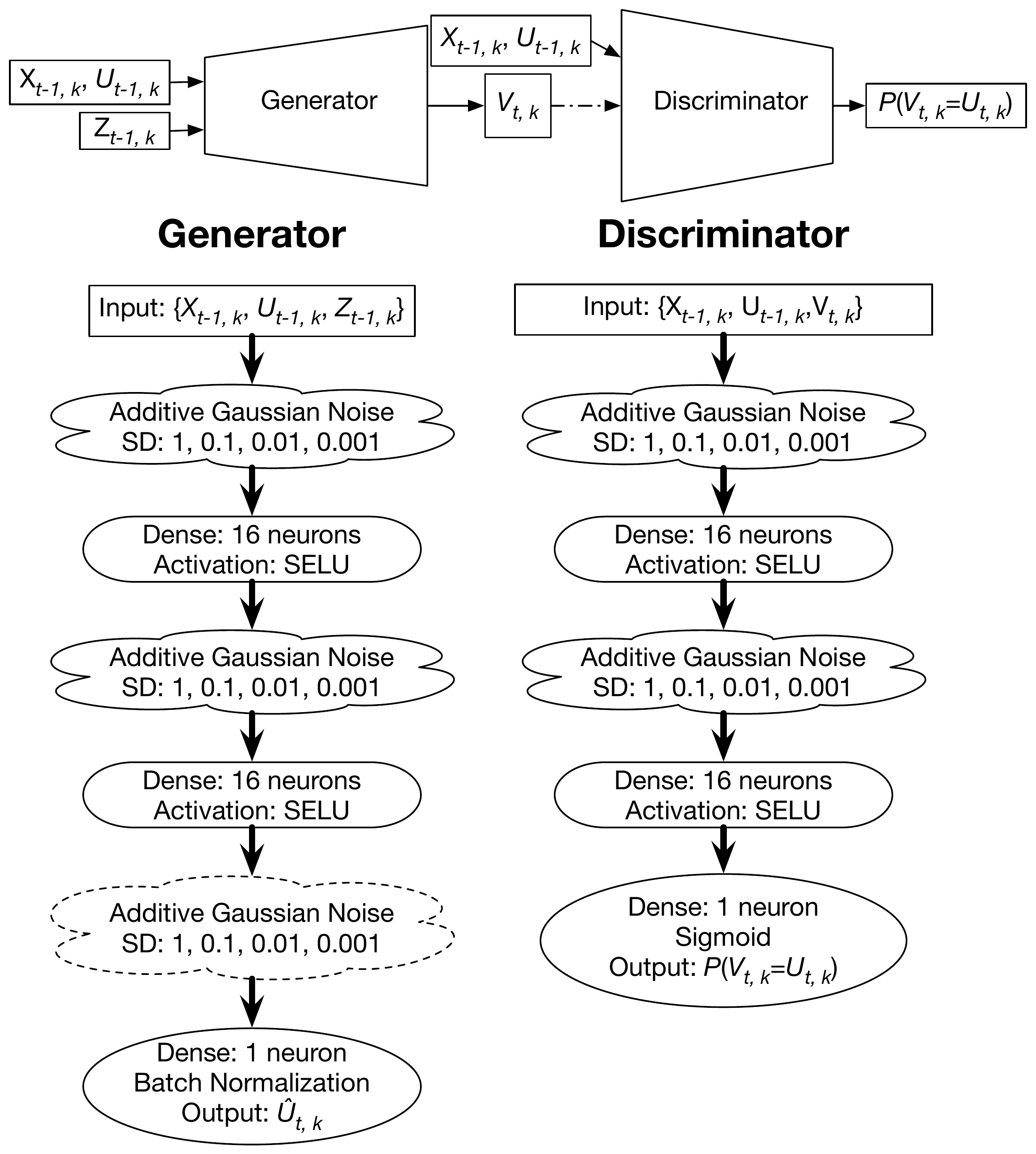}
\caption{(Top) A diagram of how the GAN networks are connected for training. (Bottom) A diagram of the GAN network architectures used for the stochastic parameterization.}
\label{gan_diagram}
\end{figure}

Each GAN we consider consists of the same neural network architecture with variations in the inputs and how noise is scaled and inserted into the network. A diagram of the architecture of the GAN networks is shown in Fig. \ref{gan_diagram}. Both the generator and discriminator networks contain two hidden layers with 16 neurons in each layer. The weights of the hidden layers are regularized with a L2, or Ridge, penalty \cite{Hoerl1970} with scale factor $\lambda$ of 0.001. Scaled exponential linear unit (SELU) activation functions \cite{Klambauer2017} follow each hidden layer. SELU is a variation of the common Rectified Linear Unit (ReLU) activation function with a scaled exponential transform for the negative values that helps ensure the output distribution retains a mean of 0 and standard deviation of 1.  Larger numbers of neurons per hidden layer were evaluated and did not result in improved performance. Gaussian additive noise layers before each hidden layer and optionally the output layer inject noise into the latent representations of the input data. A batch normalization \cite{Ioffe2015} output layer ensures that the output values retain a mean of 0 and and standard deviation of 1, which helps the generator converge to the true distribution faster.

The GAN training procedure iteratively updates the discriminator and generator networks until the networks reach an adversarial equilibrium in which the discriminator should not be able to distinguish ``true'' data from generator samples. The inputs, outputs, and connections between networks are shown in Fig. \ref{gan_diagram}. A batch, or random subset of samples drawn without replacement from the training data, of truth run output is split in half and one subset is fed through the generator and then into the discriminator or is sent directly to the discriminator. The discriminator is then updated based on errors in predicting which samples have passed through the generator. Another batch of samples are drawn and sent through the generator and then to a connected discriminator with frozen weights. The gradient with respect to the generator is calculated based on the reversed labeling that the generated samples originate from the true training data. This reversal forces the discriminator to teach the generator features that would worsen the discriminator's own performance. Gaussian noise is injected into the neural network at each iteration through both the input and hidden layers. We consider hidden layer Gaussian noise scaled to standard deviations at different orders of magnitude in order to determine how the magnitude of the noise affects the forecast spread and the representation of the model climate. In the training process, neural network weights are updated based on subsamples of examples drawn randomly from the training data without regard for temporal dependence. The GAN is trained as a map between $X_{t-1,k}$ (and possibly $U_{t-1,k}$) and the sub-grid-scale tendency $U_{t,k}$.  In forecast mode, we test providing both white, or uncorrelated noise, and red, or correlated noise to the GAN. The red noise is generated using an AR(1) process with a correlation equal to the lag-1 autocorrelation of the deterministic residuals of the GAN. The color of the noise is not relevant during the training process: both white- and red-noise representations are trained in the same way. The noise values are kept constant through the integration of a single timestep. The difference between the white- and red-noise representations only manifests when they are incorporated as parameterizations in the full model (Eqn. \ref{eq:L96}).

The GANs are all trained with a consistent set of optimization parameters. The GANs are updated through stochastic gradient descent with a batch size (number of examples randomly drawn without replacement from the training data) of 1024 and a learning rate of 0.0001 with the Adam optimizer \cite{Kingma2014}. The GANs are trained for 30 epochs, or passes through the training data. The model weights are saved for analysis every epoch for the first 20 epochs and then every 2 epochs between epochs 20 and 30. The GANs are developed with the Keras v2.2 machine learning API coupled with Tensorflow v1.13.

The GAN configurations considered in this study are summarized in Table~\ref{tab:gan_configs}. A short name of the format `predictors--noise magnitude--noise correlation' is introduced to simplify identification of different GANs. For example, `XU-med-r' refers to the GAN that takes X and U as predictors, and uses medium (med) magnitude red (r) noise. While most GANs tested include the optional additive noise layer before the output layer, the sensitivity to this choice was also considered. GANs that do not include this final noise layer follow the naming convention above, but are indicated by an asterisk.

\begin{table}
 \caption{\label{tab:gan_configs}Summary of the GAN configurations tested.}
 \centering
 \begin{tabular}{l r r r r}
 \hline
  Short name  & Input variables & Noise magnitude & Noise correlation & Output layer noise?\\
 \hline
   XU-lrg-w  & $X_{t-1, k}$, $U_{t-1, k}$ & 1        & white & yes\\   
   XU-med-w  & $X_{t-1, k}$, $U_{t-1, k}$ & 0.1      & white & yes\\   
   XU-sml-w  & $X_{t-1, k}$, $U_{t-1, k}$ & 0.01     & white & yes\\   
   XU-tny-w  & $X_{t-1, k}$, $U_{t-1, k}$ & 0.001    & white & yes\\   
   X-med-w   & $X_{t-1, k}$            & 0.1     & white & yes\\   
   X-sml-w   & $X_{t-1, k}$            & 0.01     & white & yes\\   
   X-tny-w   & $X_{t-1, k}$            & 0.001    & white & yes\\   
   XU-lrg-r  & $X_{t-1, k}$, $U_{t-1, k}$ & 1        & red   & yes\\   
   XU-med-r  & $X_{t-1, k}$, $U_{t-1, k}$ & 0.1      & red   & yes\\   
   XU-sml-r  & $X_{t-1, k}$, $U_{t-1, k}$ & 0.01     & red  & yes \\   
   XU-tny-r  & $X_{t-1, k}$, $U_{t-1, k}$ & 0.001    & red  & yes \\   
   X-med-w   & $X_{t-1, k}$            & 0.1     & red & yes\\   
   X-sml-r   & $X_{t-1, k}$            & 0.01     & red  & yes \\   
   X-tny-r   & $X_{t-1, k}$            & 0.001    & red  & yes \\   
   XU-lrg-w*  & $X_{t-1, k}$, $U_{t-1, k}$ & 1        & white & no\\   
   XU-med-w*  & $X_{t-1, k}$, $U_{t-1, k}$ & 0.1      & white & no\\   
   XU-sml-w*  & $X_{t-1, k}$, $U_{t-1, k}$ & 0.01     & white & no\\   
   XU-tny-w*  & $X_{t-1, k}$, $U_{t-1, k}$ & 0.001    & white & no\\   
   X-sml-w*   & $X_{t-1, k}$            & 0.01     & white & no\\   
   X-tny-w*   & $X_{t-1, k}$            & 0.001    & white & no\\   
 \hline
 \end{tabular}
\end{table}

\subsection{Polynomial Regression Parameterization}

The GAN stochastic parameterization is evaluated against a cubic polynomial regression parameterization, $\hat{U}_{t,k}$, similar to the model used in \citeA{Arnold2013}. 
\begin{align}
\hat{U}_{t,k}&=U^d_{t,k} + \epsilon_{t,k} \nonumber \\
U^d_{t,k}&=aX_{t-1,k}^3+bX_{t-1,k}^2+cX_{t-1,k} + d 
\label{eq:U}
\end{align}
The parameters $[a,b,c,d]$ are determined by a least squares fit to the $(X,U)$ data from the L96 ``truth'' training run. It is known that the simple cubic polynomial deterministic parameterization $U^d_{t,k}$ is a poor forecast model for the L96 system \cite{wilks2005,Arnold2013,cmp15}, as $X$ does not uniquely determine $U$. The variability in the $(X,U)$ relationship is accounted for using a temporally correlated additive noise term:
\begin{align}
\epsilon_{t,k} &= \phi \epsilon_{t-1,k} + \sigma_\epsilon (1-\phi^2)^{1/2} z_{t,k}, \label{eq:AR1}
\end{align}
where $z\sim\mathcal{N}(0,1)$, the first order autoregressive parameters $(\phi,\sigma_\epsilon)$ are estimated from the residual process $r_t = U_t - U^d_t$, and the $\epsilon_k$ processes are independent for different $X$ variables. 

The polynomial parameterization has been specifically designed to represent the impact of the $Y$ variables in this version of the L96 model, just as traditional parameterization schemes are designed to represent a given process in a GCM. Previous studies have demonstrated that the polynomial parameterization with additive noise performs very well \cite{wilks2005,Arnold2013,cmp15}. This `bespoke' parameterization is therefore a stringent benchmark against which to test GAN parameterizations.

\section{Results}

\subsection{Metrics}

The accuracy of ensemble weather forecasts can be summarized by evaluating the root mean square error (RMSE) of the ensemble mean. The lower the RMSE, the more accurate the forecast. The RMSE at lead time $\tau$ is defined as:

\begin{equation}
RMSE(\tau) = {\sqrt{ \frac{1}{N} \sum\limits_{t=1}^N\left(X^o(t) - X^m(t_{\mathrm{init}},\tau) \right)^2}}\hspace{1mm},
\end{equation}
where $N$ is the number of forecast-observation pairs, $X^o(t)$ is the observed state at time $t$, and $X^m(t_{\mathrm{init}},\tau)$ is the ensemble mean forecast at lead time $\tau$, initialized at $t_{\mathrm{init}}$, such that $t = t_{\mathrm{init}}+\tau$.

If an ensemble forecast correctly accounts for all sources of uncertainty such that the forecast of the spread of the ensemble and measured probabilities of an event are statistically consistent, the forecast is said to be \emph{reliable} \cite{Wilks2011}. In this study we assess the reliability of the ensemble using the spread-error relationship \cite{leutbecher2008,leutbecher2010}. This states that, for an unbiased and reliable forecasting system, the root mean square error in the ensemble mean is related to the average ensemble variance:
\begin{equation}
\frac{M}{M-1} \; \overline{\text{estimate ensemble variance}} = \frac{M}{M+1} \; \overline{\text{mean square error}}, \label{eq:variancespread}
\end{equation}
where $M$ is the number of ensemble members, and the variance and mean error have been estimated by averaging over many forecast-verification pairs. For the large ensemble size used in this study, $M=40$, we can consider the correction factor to be close to 1. A skilful probabilistic forecast will have as small an RMSE as possible, while also demonstrating a statistical match between RMSE and average ensemble spread.

The simplest definition of the `climate' of the L96 system is the probability density function (PDF) of the individual $X_{t,k}$ values. The climatological skill can therefore be summarized by quantifying the difference between the true and forecast PDF. The Hellinger distance $H$, is calculated for each forecast model:
\begin{equation}
H(p,q) = \frac{1}{2} \int \left( \sqrt{p(x)} - \sqrt{q(x)} \right)^2 \mathrm{d}x,
\end{equation}
where $p(x)$ is the forecast PDF, and $q(x)$ is the verification PDF \cite{pollard2002}. The smaller $H$, the closer the forecast pdf is to the true pdf. We also considered the Kullback-Leibler ($KL$) divergence \cite{kullback1951}, motivated by information theory, but found it provided no additional information over the Hellinger distance, so results for the $KL$ are not shown for brevity. 

Evidence that the L96 model displays distinct dynamical regimes of behavior for the parameter set considered was presented in \cite{cmp15}, in which regime affiliations were determined using a metric based on the temporally local covariance. \cite{cmp15} found that during the more common regime (regime frequency $\sim$ 80\%), the eight $X$ variables exhibit wave-2 like behavior around the ring, while in the rarer regime, the $X$ variables exhibit wave-1 type behavior. Another approach to characterizing regime structure that makes use of both the instantaneous state and the recent past of the system is Hidden Markov Model (HMM) analysis \cite{rabiner89,fcfm08,mrhm15}. In an HMM analysis, it is assumed that underlying the observed state variables is an unobserved Markov chain taking discrete values.  The HMM algorithm provides maximum likelihood estimates of the probability distributions of the state variables conditioned on the instantaneous hidden state values, the stochastic matrix $Q$ of transition probabilities for each time step, and an optimal estimate of the hidden state sequence.

\subsection{Offline assessment of GAN performance}

\begin{figure}
    \centering
    \includegraphics[width=\textwidth]{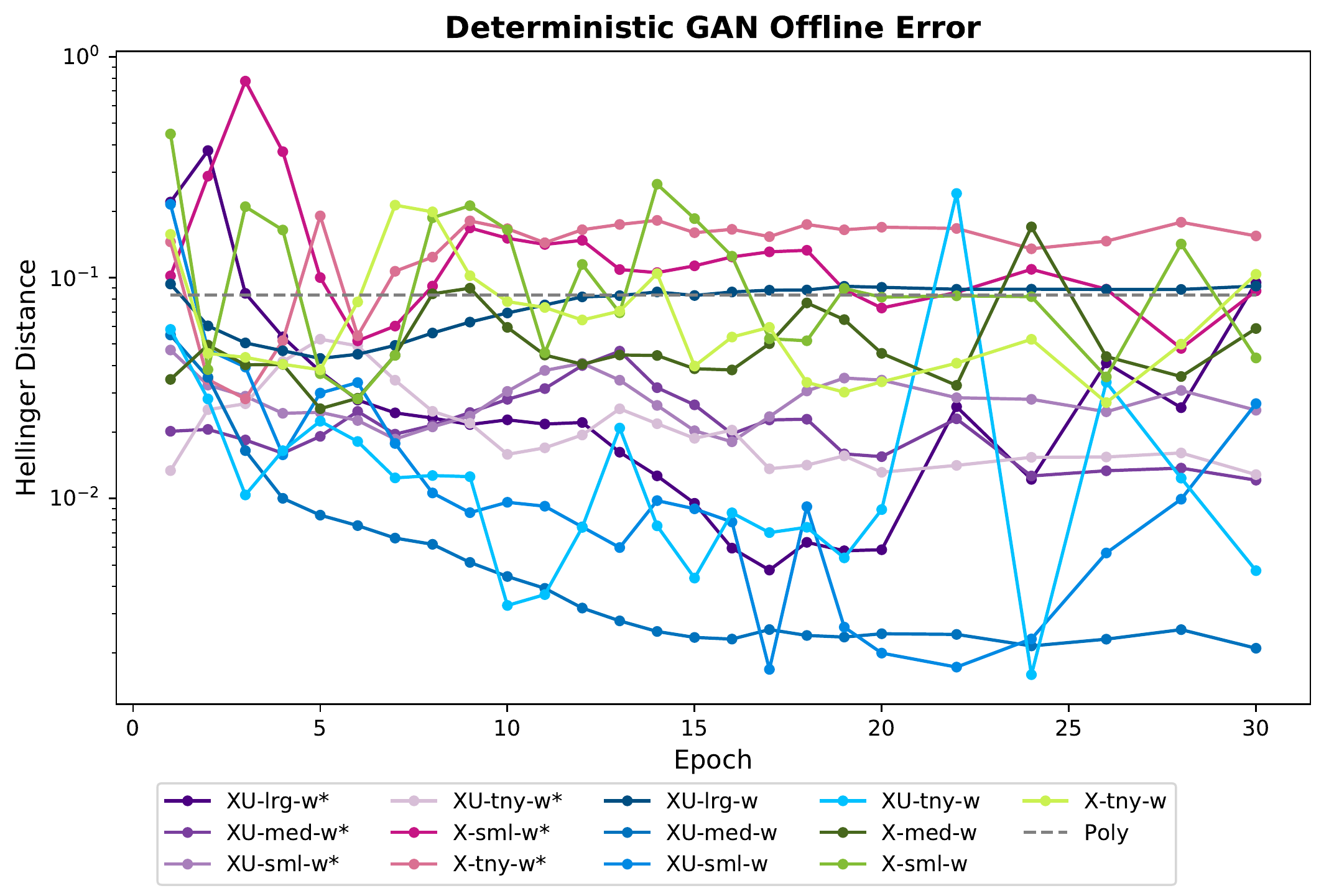}
    \caption{Offline assessment of GAN performance. Hellinger distances between the GAN sub-grid tendency distributions given input $X$ and $U$ values from the truth run, and the truth run sub-grid forcing distribution as a function of training epoch.}
    \label{fig:gan_offline}
\end{figure}
The GAN parameterizations are first evaluated on how closely their output sub-grid forcing distributions match those of the truth run when the GANs are supplied with input $X$ and $U$ values from the truth run. This is  summarized by the Hellinger distance in Figure~\ref{fig:gan_offline}. Most of the GANs show a trend of decreasing Hellinger distance for the first few epochs followed by mostly stable oscillations. GANs with both $X_{t-1,k}$ and $U_{t-1,k}$ as input tend to perform better in the offline analysis than those with only $X_{t-1,k}$. Larger input noise standard deviations seem to reduce the amount of fluctuation in the Hellinger distance between epochs, but there does not appear to be a consistent correlation with noise standard deviation and Hellinger distance. Note that the weights as fitted at the end of epoch 30 are used in the forecast networks, regardless of whether the GAN at this epoch shows the minimum Hellinger distance.

\subsection{GAN simulation of sub-grid-scale tendency distribution}

\begin{figure}
    \centering
    \includegraphics[width=\textwidth]{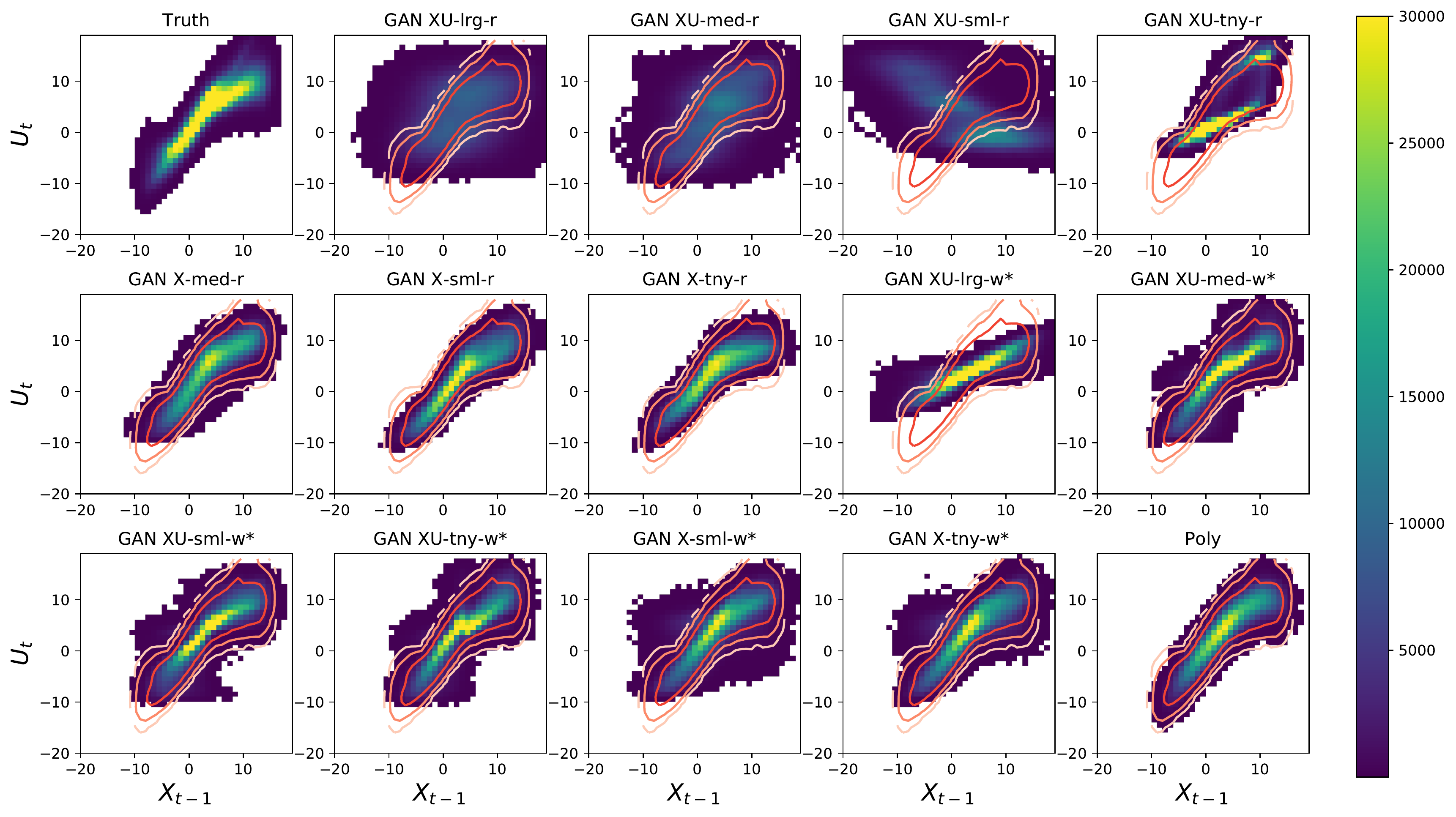}
    \caption{Joint distributions (2D histograms) of $X_{t-1}$ and $U_{t}$ for each GAN configuration. The truth joint distribution is overlaid in red contours on each forecast model distribution.}
    \label{fig:climate_2d_hist}
\end{figure}

The joint distributions of $X_{t-1}$ and $U_t$ from the  different model runs reveal how the noise standard deviation affects the model climate (Fig. \ref{fig:climate_2d_hist}). Larger noise standard deviations increase the range of $X$ values appearing in the run but do not appear to change the range of $U$ values output by the GAN. The behavior of the XU-tny-r GAN devolved into oscillating between two extremes. The X-only GANs did the best job in capturing the shape of the truth distribution although they underestimated the variance at the extremes. While the XU-w* and X-w* series captured the bulk of the distribution well, there were spurious points outside the bounds of the truth distribution for all of these models. The Polynomial model captured the conditional mean and variance of the distribution well but smoothed over some of the subtle details of the joint distribution that some of the GANs were able to capture.

\subsection{Weather Evaluation}
\begin{figure}
\centering
\includegraphics[width=0.95\textwidth]{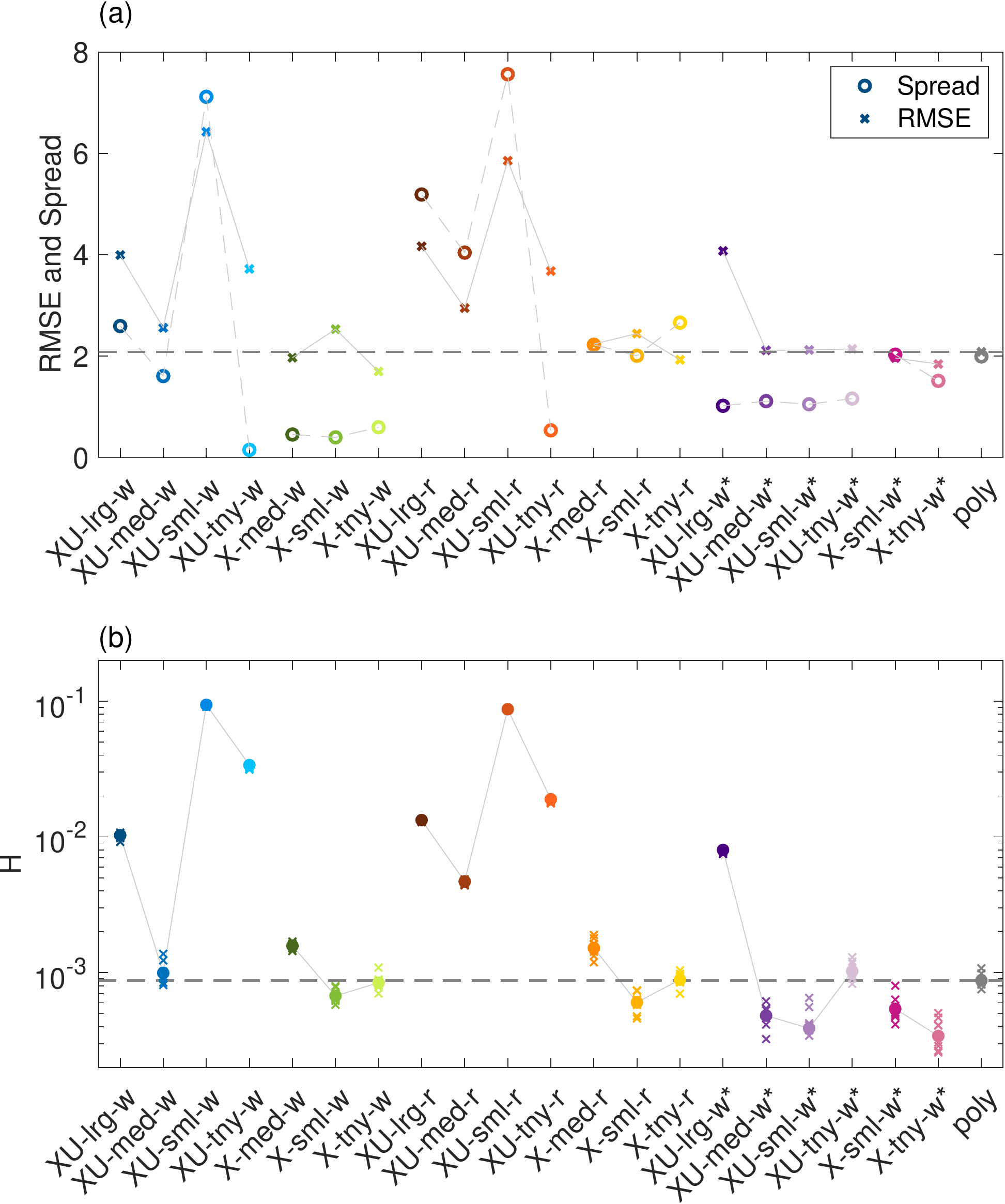}
\caption{Summary of performance of different parameterized models (x-axis) (a) Weather forecast performance. Ensemble spread (circles) and RMSE (crosses) for experiments with white and red noise in GANs at timestep 201. The horizontal dashed line indicates the RMSE for the polynomial forecast model. Ideally, a forecast model will produce forecasts with small RMSE while maintaining the match between spread and RMSE. (b) Climate performance. The Hellinger distance between each forecast pdf and the `true' pdf. The metric in question is calculated for the best estimate of the climatological pdf, averaging across all $X$ variables (circles), as well as for each X variable in turn, e.g. comparing true and forecast $X_1$ pdfs, etc (crosses). The latter gives an indication of the sampling uncertainty. The horizontal dashed line indicates the mean value of $H$ for the polynomial model.}
\label{fig:weath_clim_metrics}
\end{figure}

\begin{figure}
\centering
 \includegraphics[width=\textwidth]{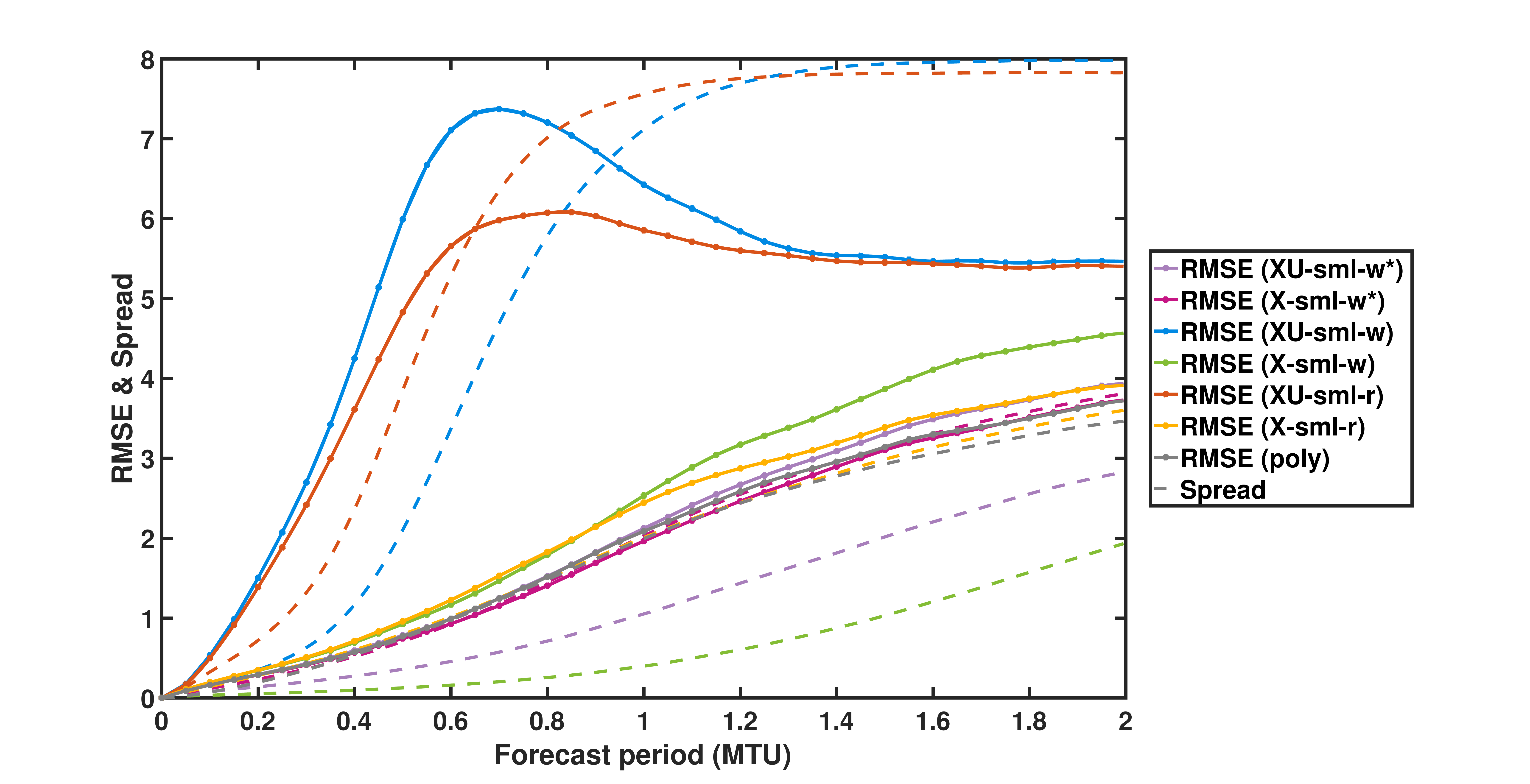}
\caption{RMSE (lines with dots) and ensemble spread (dashed lines) for a subset of experiments with white and red noise in GANs. Note that 400 forecast time steps corresponds to 2 MTU, or 10 `atmospheric days'.}
\label{fig:gan_rmse}
\end{figure}

The parameterized models for the Lorenz '96 system are evaluated in a weather forecast framework. An extensive set of re-forecast experiments were produced for 751 initial conditions selected from the attractor. An ensemble of 40 forecasts was produced from each initial condition (i.e. no initial condition perturbations are used). Different random seeds are used for each ensemble member to generate the stochastic perturbations used in the GAN or polynomial parameterizations.

Figure \ref{fig:weath_clim_metrics} shows the RMSE and spread for all weather experiments at 1 MTU. A reduction in RMSE indicates an ensemble forecast that more closely tracks the observations. A good match between RMSE and ensemble spread indicates a reliable forecast. The best performing GANs in terms of RMSE are X-tny-r, X-tny-w, and X-tny-w*. All of these models performed slightly better than the polynomial regression, which was competitive with most GANs in terms of both RMSE and the ratio of RMSE to spread. The spread of the white noise GANs is generally underdispersive. Most of the red noise GANs, on the other hand, are somewhat overdispersive with X-med-r having the spread/error ratio closest to 1. Figure \ref{fig:gan_rmse} shows the RMSE and ensemble spread for a subset of the ensemble forecasts of the $X$ variables performed using the GAN parameterizations or the bespoke polynomial parameterization. X-sml-w* demonstrates both low RMSE and similar spread to RMSE throughout the forecast period. X-sml-r and X-sml-w feature similar RMSE through 1 MTU, but X-sml-r has smaller RMSE after that point and a better spread-to-error ratio throughout the period. The XU GANs have higher RMSEs than their X counterparts because the XU models may have overfit to the strong correlation between $U_{t-1,k}$ and $U_{t,k}$ in the training data. Inspection of the input weights revealed that XU GANs generally weigh $U_{t-1,k}$ more highly than $X_{t-1,k}$. At maxima and minima in the waves, the XU models may be biased toward extending the current growth forward, which can be a source of error in forecast runs.  


\subsection{Climate Evaluation}

\begin{figure}
   \centering
   \includegraphics[width=1.0\textwidth]{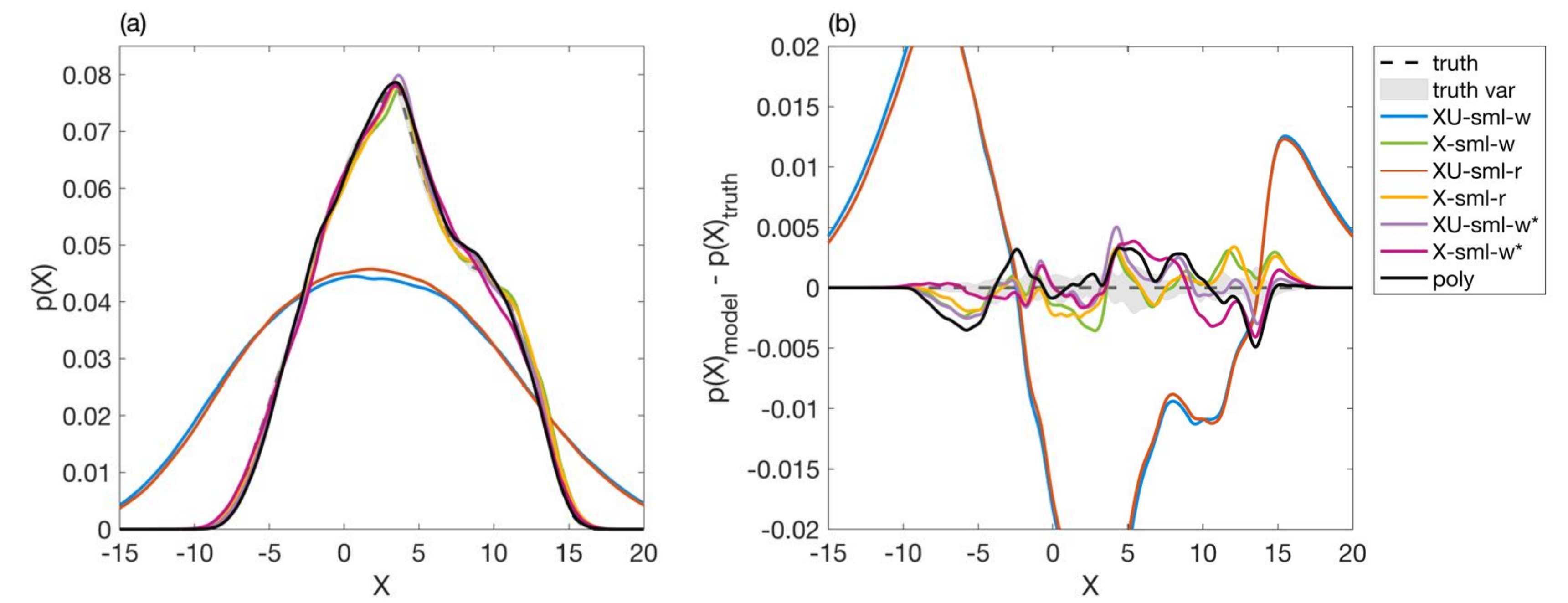}
   \caption{The skill of the forecast models at reproducing the climate of the Lorenz `96 system defined as the pdf of the $X$ variables. (a) The pdf of the Lorenz `96 system (black dashed) compared to a subset of the forecast models. (b) The difference between the forecast and true pdfs shown in (a). Sampling variability is indicated by shading the range in metrics for each of the 8 $X$ variables of the full Lorenz `96 system.}
   \label{fig:clim_pdf}
\end{figure}

The GAN parameterizations are also tested on their ability to characterize the climate of the Lorenz `96 system. First, the ability to reproduce the pdf of the $X$ variables is evaluated. Each forecast model  and the full L96 system were used to produce a long simulation of length 10,000 MTU. Figure~\ref{fig:clim_pdf}(a) shows  kernel density estimates of the marginal pdfs of $X_{t,k}$ from the full L96 system and from a sample of parameterized models. The pdf of the true L96 system is markedly non-Gaussian, with enhanced density forming a `hump' at around $X=8$. Compared to the true distribution, the XU-sml-w and XU-sml-r models both perform poorly, producing very similar pdfs with too large a standard deviation and that are too symmetric. However, the other models shown skilfully reproduce the true pdf. Figure ~\ref{fig:clim_pdf}(b) shows the difference between each forecast pdf and the true pdf. Several GANs perform as well if not better than the benchmark bespoke polynomial parameterization. 

Figure~\ref{fig:weath_clim_metrics}(b) shows the Hellinger distance $H$ evaluated for each parameterized model. The filled circles indicate the value of the metric when the pdf is evaluated across all $X$ variables in both parameterized and truth timeseries. The crosses give an indication of sampling variability, and indicate the metrics comparing pairs of $X$ variables, i.e. $X_{t,j}$ and $X_{t,k}$ for $j\neq k$. Quantifying the parameterized model performance in this way allows for easy ranking of the different models. While the AR(1) stochastic polynomial parameterized forecast model is very skillful \cite{Arnold2013}, several GANs outperform the polynomial model.

\begin{figure}
   \centering
   \includegraphics[width=1.0\textwidth]{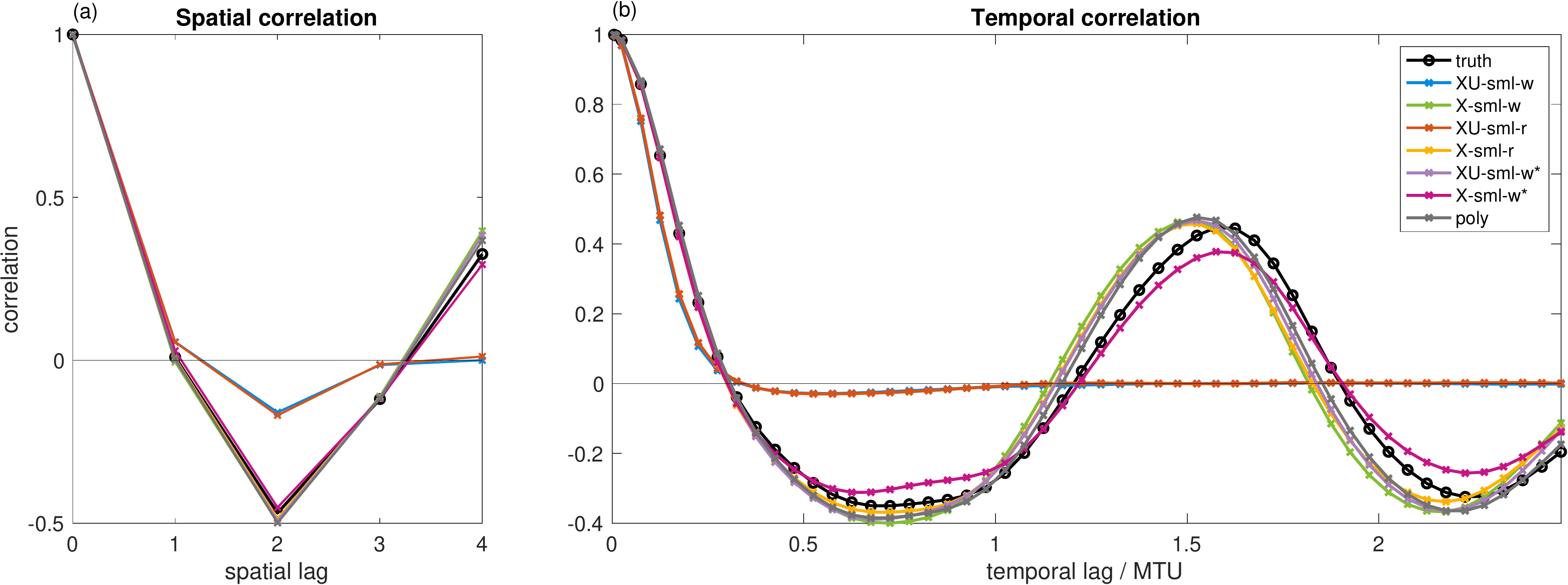}
   \caption{The skill of the parameterized models (colors) at reproducing the `true' (a) spatial correlation and (b) temporal correlation of the $X$ variables in the Lorenz `96 system (black), calculated from the climatological simulation. The sampling variability in these metrics, as indicated by the variability between the metric calculated for different $X$ variables, is narrower than the plotted line width.}
   \label{fig:clim_st_corr}
\end{figure}

In addition to correctly capturing the distribution of the $X$ variables, it is desirable that a  parameterized model will capture the spatio-temporal behavior of the system. This is assessed by considering the spatial and temporal correlations in both the true system and parameterized models. The diagnostic is shown for a subset of the tested paramererized models in Figure~\ref{fig:clim_st_corr}. It is evident that the parameterized models that skillfully capture the pdf of $X$ also skillfully represent the spatio-temporal characteristics of the system. The X-sml-w* scheme performs particularly well, improving over the stochastic polynomial approach. 

\begin{figure}
\centering
\includegraphics[width=1.0\textwidth]{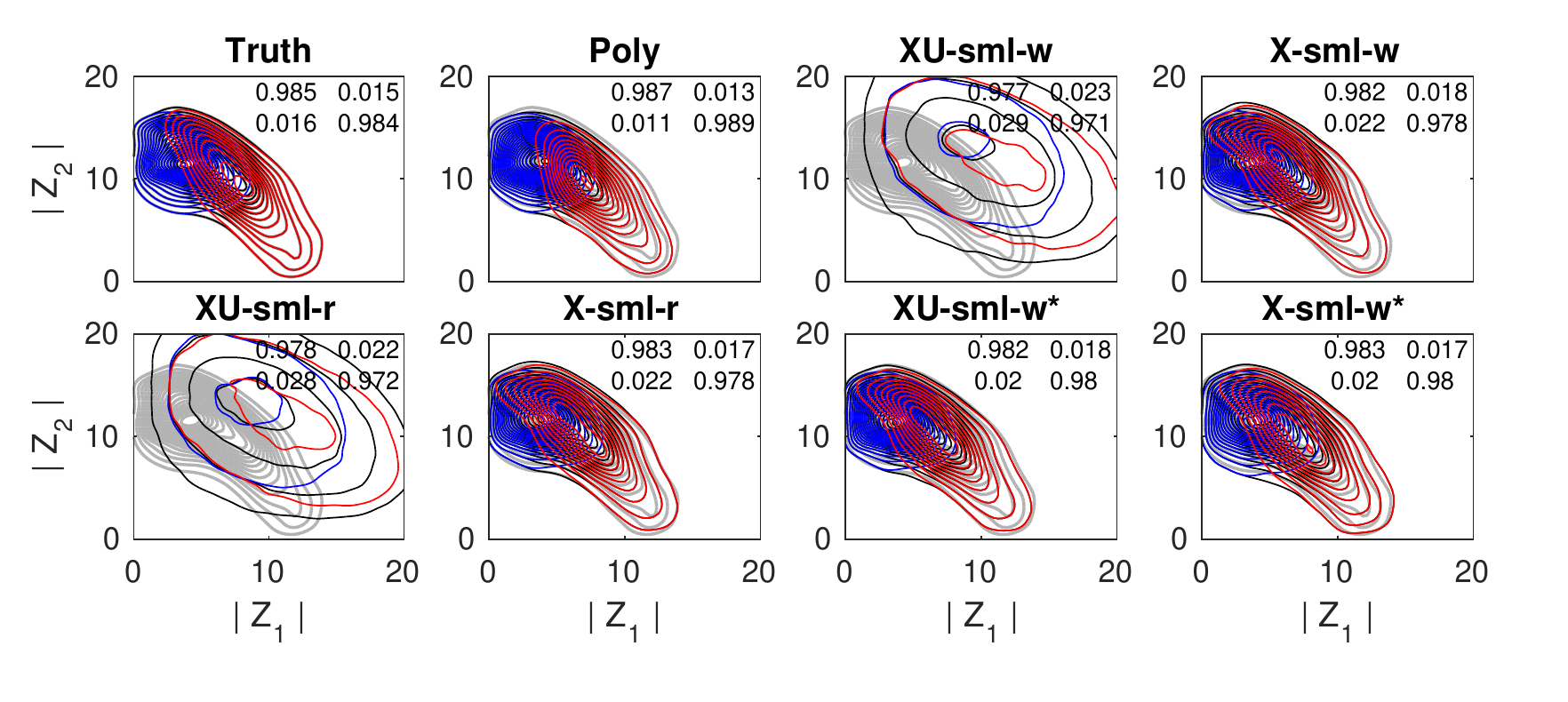}
\caption{Joint distributions of the projections of $X$ on wavenumbers 1 and 2 ($|Z_1|$ and $|Z_2|$ respectively) conditioned on HMM regime occupation (red and blue contours) for the truth simulation subset of parameterized simulations.  In all but the upper left panel, thin black lines display the unconditional joint distribution.  In all panels the grey curves denote the full joint distribution (without regime occupation conditioning) from the truth simulation.  The conditional distributions have been scaled by the relative probabilities of each state. Inset: HMM stochastic matrix $Q$. }
\label{fig:hmm_joint_distributions}
\end{figure}

Following the regime results presented in \cite{cmp15}, we use HMM analysis to classify into two clusters the instantaneous states in the 4-dimensional space spanned by the norms of the projections of $X$ on wavenumbers one through four (denoted $Z_{j}, j=1,...,4$).  Because of the spatial homogeneity of the $X_{t,k}$ statistics, these wavenumber projections correspond to the Empirical Orthogonal Functions.

Kernel density estimates of the joint pdfs of the projections of $X$ on wavenumbers 1 and 2 are presented in Figure \ref{fig:hmm_joint_distributions}, along with estimates of the joint distributions conditioned on the HMM regime sequence. The conditional distributions have been scaled by the probability of each state so that the full joint pdf is the sum of the conditional pdfs. For reference, the unconditional joint pdf for truth is shown in gray contours in each panel. The stochastic matrix $Q$ shows the probability of remaining in each regime (diagonal values) or transitioning from one regime to the other (off-diagonal values). As in Figures \ref{fig:clim_pdf} and \ref{fig:clim_st_corr}, only a subset of results are displayed. 

The clear separation of the truth simulation into two distinct regimes is modeled well by the polynomial parameterization and most of the GAN parameterizations.  With the exception of XU-sml-w and XU-sml-r, the regime spatial structures and stochastic matrices are captured well. The GANs are slightly more likely to transition between regimes than the truth run, while the polynomial run is slightly more likely to stay in the same regime. Consistent with the other climate performance results presented above, the joint distribution of $|Z_{i}|$ produced by XU-sml-w and XU-sml-r are strongly biased, with no evidence of a meaningful separation into two distinct regimes of behavior. 

\begin{figure}
   \centering
   \includegraphics[width=0.8\textwidth]{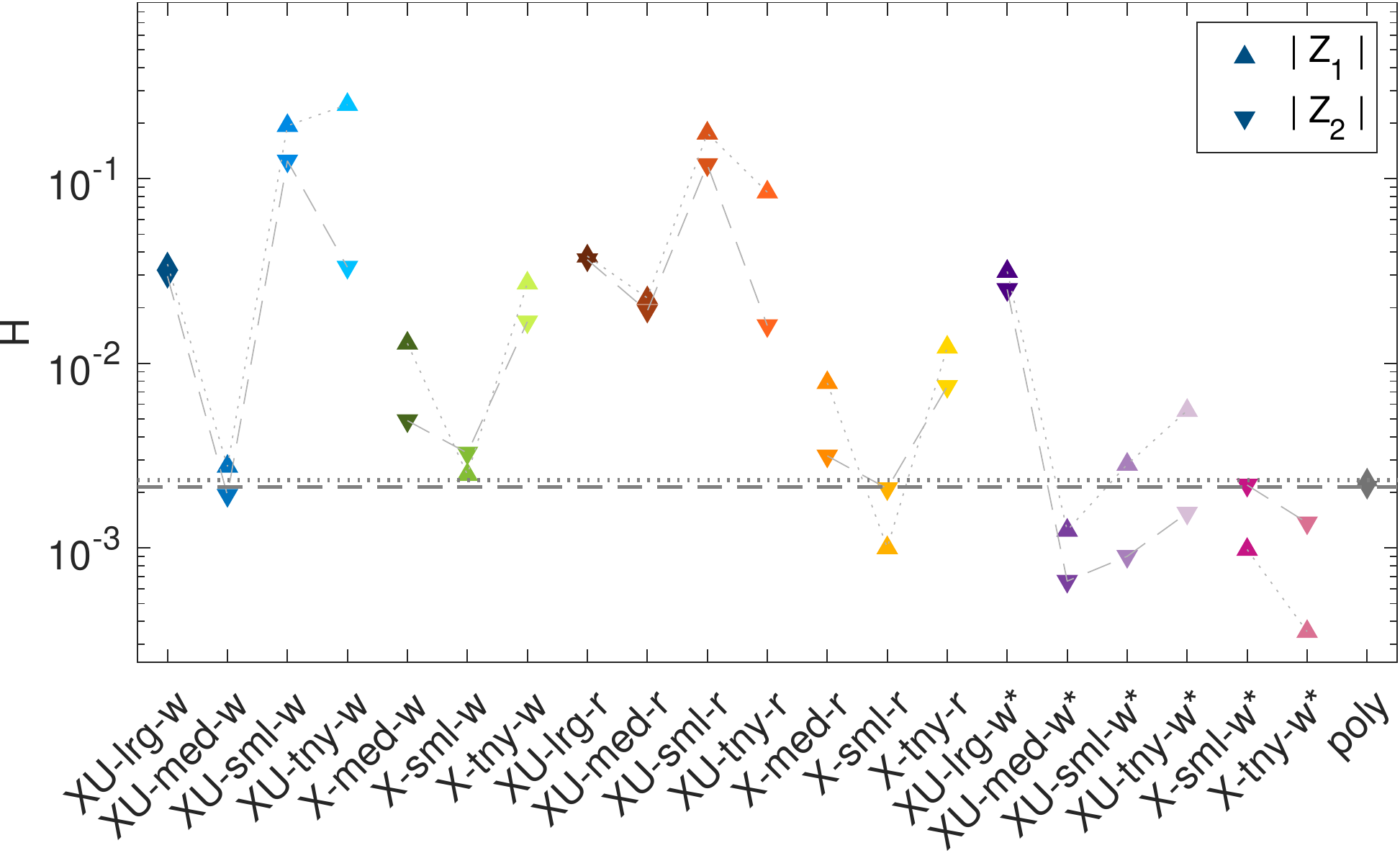}
   \caption{The Hellinger distance between each forecast pdf and the `true' pdf, considering the projection of the $X$ variables onto (upward triangles) wavenumber 1, and (upward triangles) wavenumber 2.}
   \label{fig:clim_eof_hell}
\end{figure}

To quantify the forecast model skill at reproducing the L96 behavior in wavenumber space, Figure~\ref{fig:clim_eof_hell} shows $H$ calculated over the marginal pdfs of $|Z_{1}|$ and $|Z_{2}|$ as upward and downward triangles respectively. Several of the GAN evaluated are competitive with, or improve upon, the polynomial parameterization scheme. The best performing GANs also performed the best in terms of other climate metrics (e.g. see Figure~\ref{fig:weath_clim_metrics}).

We note that in particular, models which accurately capture the regime behavior will also show good correlation statistics when averaged over a long time series. The regime analysis can help diagnose why a model shows poor correlation statistics. For example, X-sml-w* accurately captures the marginal distributions of the two regimes, but the frequency of occurrence of the red regime, dominated by wavenumber-1 behavior, is slightly too high (at 54\% compared to 51\% in the `truth' run). This reduces the magnitude of the negative (positive) correlation at a lag of 0.75 (1.5) MTU observed in figure~\ref{fig:weath_clim_corr}.

\subsection{Wavelet Analysis}
\begin{figure}
    \centering
    \includegraphics[width=\textwidth]{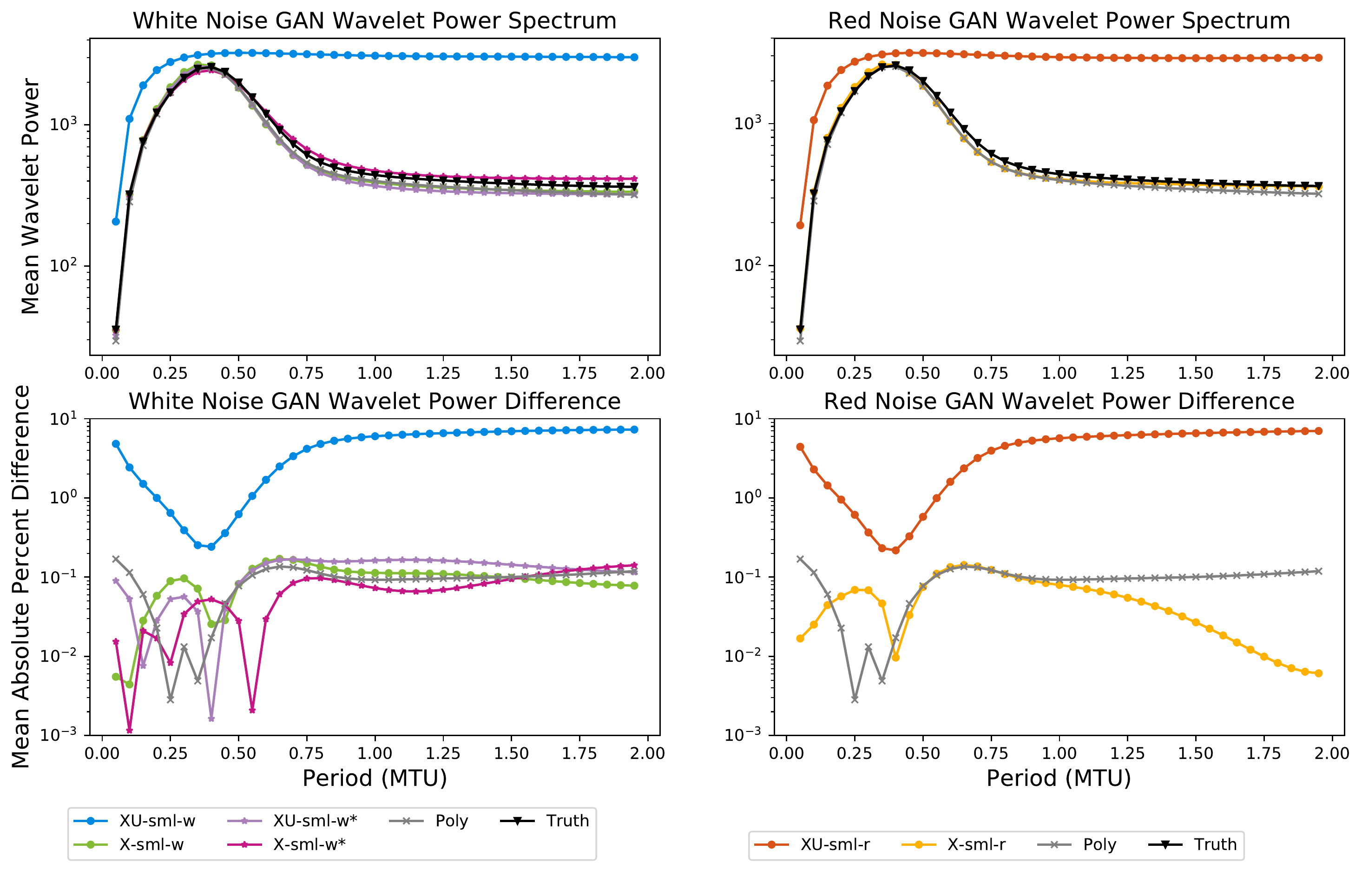}
    \caption{(Top) Wavelet power spectra for the white and red noise GAN climate runs as well as the polynomial and truth runs. (Bottom) Mean absolute percent differences between the truth wavelet power spectrum and each model.}
    \label{fig:gan_climate_wavelet}
\end{figure}

To further investigate why the white noise and red noise GANs differ in operational performance, a wavelet analysis is performed on time series of outputs from the climate runs. A discrete wavelet transform decomposes the time series into contributions from different periods. The total energy $E$ for a given period is represented as 
\begin{equation}
    E=\frac{1}{T}\sum_{t=1}^{T}w_t^2
\end{equation}
where $w$ is the wavelet magnitude at a given timestep $t$. The total power for each period from each model is shown in Fig. \ref{fig:gan_climate_wavelet}. All sml GANs except the XU-sml-r and XU-sml-w follow the truth power curve closely. follow The polynomial regression follows the truth closely although it tends to underestimate the power slightly for each period. The GANs peak in power at the same period when the temporal correlation in Fig. \ref{fig:clim_st_corr} crosses 0. The GANs with poor Hellinger distances also contained more energy for longer periods. 

A clearer evaluation of the wavelet differences can be found by calculating the mean absolute percentage difference from the truth run at different wavelengths. The absolute difference between the truth and parameterized runs increases with increasing wavelengths, so the percentage difference is employed to control for this trend:
\begin{equation}
    MAPD=\frac{1}{T}\sum_{t=1}^{T}\frac{|E_{g,t}-E_{u,t}|}{E_{u,t}}
\end{equation}
The MAPD scores with wavelength in Fig. \ref{fig:gan_climate_wavelet} shows that none of the GANs consistently perform better at all periods, but some do provide closer matches to the truth spectrum for the longer periods. In the peak energy period, the different GANs have minimum error for slightly different periods before increasing in error again. The X-sml-r GAN uniquely shows decreasing MAPD with increasing period, while the white noise GANs generally have similar differences across the range of evaluated periods.

\section{GAN Health Risks}
The disparity between the offline verification statistics and those from the climate and weather runs highlights the challenges in training GANs for parameterization tasks. Neither the values of the generator loss or the offline evaluation of the GAN samples correlated with their performance in the forecast model integrations. The generator and discriminator networks optimize until they reach an equilibrium, but there is no guarantee that the equilibrium is stable or optimal. Some of the differences in the results may be due to particular GANs converging on a poor equilibrium state as opposed to other factors being tested. GANs and other machine learning parameterization models are trained under the assumption that the data are independent and identically distributed, but in practice are applied to spatially- and temporally-correlated fields sequentially, potentially introducing nonlinear feedback effects. GANs are more complex to train than other machine learning methods because they require two neural networks and do not output an informative loss function. Larger magnitude additive noise appears to help prevent runaway feedbacks from model biases at the expense of increasing weather prediction errors. The inclusion of the batch normalization output layer appeared to assist both training and prediction by limiting the possible extremes reached during integration. 

\section{Discussion}

In this study, we chose to focus on GANs for stochastic parameterization primarily because the framework offers a way to directly embed stochasticity into the model instead of adding it \textit{a posteriori} to a deterministic parameterization. The use of the discriminator network as an adaptive loss function is also attractive because it reduces the need for developing a hand-crafted loss function and can be scaled to higher dimensional and more complex outputs.

Several of the GANs tested show a weather and climate skill that is competitive with a bespoke polynomial parameterization scheme. For climatological skill, several different metrics were considered including the distribution of the $X$ variable, spatio-temporal correlation statistics, and regime behavior. We found that forecast models that performed well according to one metric also performed consistently well for all metrics. The good performance of the GAN is encouraging, demonstrating that GANs can indeed be used as explicit stochastic parameterizations of uncertain sub-grid processes directly from data. Furthermore, a small number of GANs improve upon the bespoke polynomial approach, indicating the potential for such machine-learned approaches to improve on our current hand-designed parameterization schemes, given suitable training data. While this has been proposed and demonstrated for deterministic parameterizations \cite{Schneider2017,Rasp2018}, this is a first demonstration for the case of an explicitly stochastic parameterization.

\begin{figure}
\centering
\includegraphics[width=\textwidth]{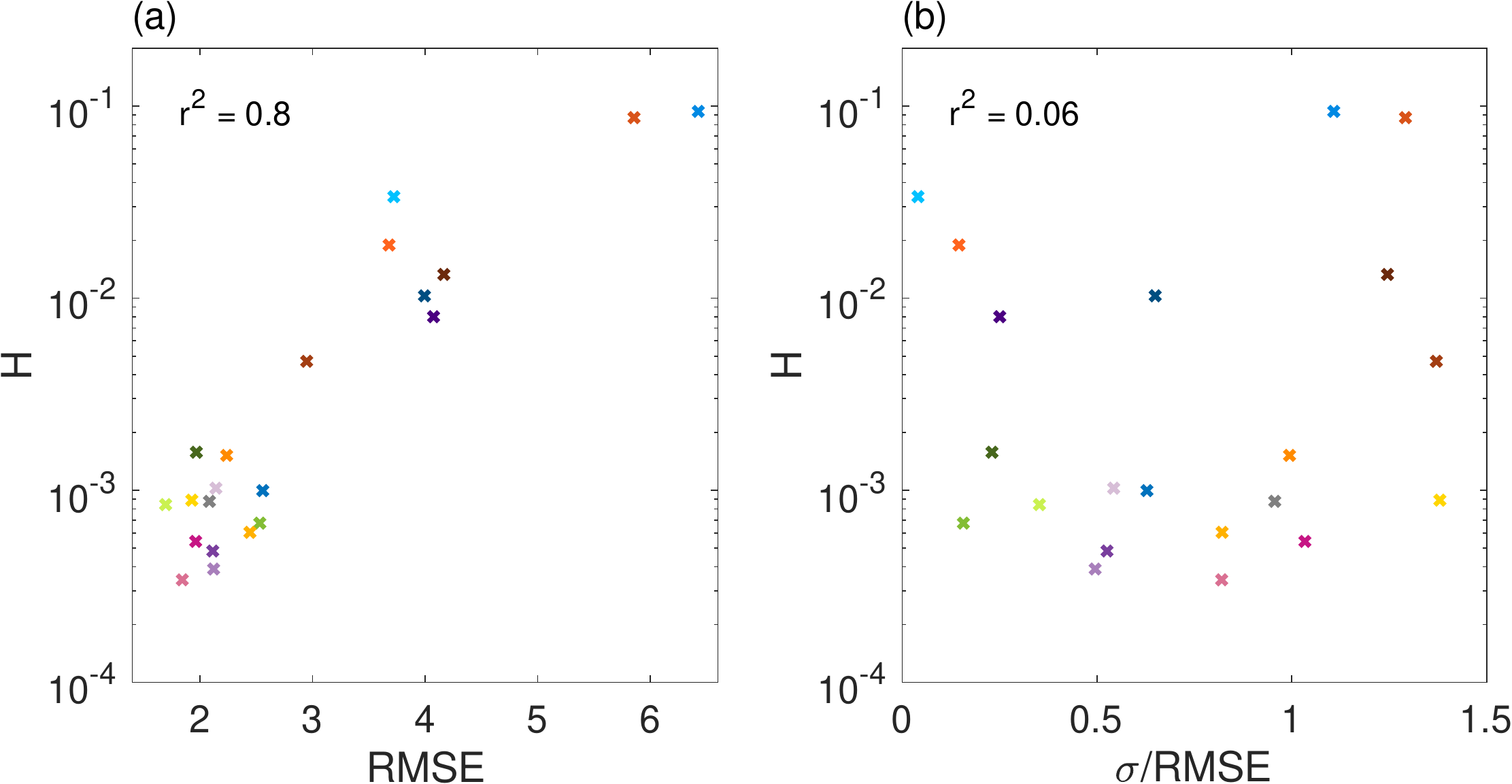}
\caption{Correlation between weather forecast skill and climate performance. (a) Weather forecast RMSE versus climate Hellinger distance. (b) Weather forecast spread-error ratio versus climate Hellinger distance. Colors indicate forecast model, as in Figure~\ref{fig:weath_clim_metrics}}
\label{fig:weath_clim_corr}
\end{figure}

Comparison of Figure~\ref{fig:weath_clim_metrics}(a) and (b) indicates a relationship between forecast models that perform well on weather and climate timescales. To quantify this further, Figure~\ref{fig:weath_clim_corr} shows the correlation between weather forecast skill and climate performance for each model considered. Models which produce weather forecasts with a lower RMSE also show good statistics on climate timescales. In contrast, the \emph{reliability} of weather forecasts, i.e., the statistical match between spread and error, is a poor predictor of climate performance. This is reflected by the competitive performance of the white noise GAN for producing a realistic climate, whereas on weather timescales, red noise increases the spread and can thereby substantially improve the forecast reliability (e.g. consider the X-med, X-sml and X-tny GAN for white and red noise respectively: Figure~\ref{fig:weath_clim_metrics}). This relationship between initialized and climatological performance has been discussed in the context of global models \cite{williams2013}. It provides further evidence that parameterizations can first be tested in weather forecasts before being used in climate models, as promoted by the `seamless prediction' framework \cite{brunet2010}.

Wavelet analysis helped uncover differences in model performance across different time scales. While the other evaluation metrics focused on distributional or error metrics in the time domain, the wavelet power spectrum separated the time series into different periods and enabled comparisons of the energy embedded in different scales. In particular, the wavelet analysis revealed that some of the GANs added energy to the system either at long periods or across all periods in some cases.

The standard deviation of the noise does impact both the training of the GANs and the resulting weather and climate model runs. Using too large a standard deviation limits the ability of the GAN to discover the structure of the true distribution of the data. Standard deviations that are too small may result in either the generator or discriminator becoming overly good at their tasks during training, which results in the GAN equilibrium being broken. During simulations, noise standard deviations that are too small can result in the system becoming trapped within one regime and never escaping.

In addition to the GAN configurations evaluated in the paper, other GAN settings were tested and were found to have similar or worse performance. Given the relative simplicity of the Lorenz '96 system, adding neurons in each hidden layer did not improve performance. Using the SELU activation function generally resulted in faster equilibrium convergence than the ReLU. Varying the scaling factor for the L2 regularization on each hidden layer did affect model performance. Using a larger value greatly reduced the variance of the predictions, but using a smaller value resulted in peaks of the final distributions being too far apart, especially when both $X_{t-1,k}$ and $U_{t-1,k}$ were used as inputs. We also tested deriving $U$ from a 1D convolutional GAN that reproduced the set of $Y$ values associated with each $X$. That approach did produce somewhat realistic $Y$ values but contained "checkerboard" artifacts from the convolutions and upscaling, especially at the edges. The sum of the $Y$ values was also not equal to $U$ derived from Eq. \ref{eq:U}.

The L96 system is commonly used as a testbed for new ideas in parameterization, and ideas tested using the system can be readily developed further for use in higher complexity Earth system models. However, the L96 system has many fewer dimensions than an Earth system model and a relatively simple target distribution. The relative simplicity of the L96 system may have also led to the more complex GAN overfitting to the data compared with the simpler polynomial parameterization. For more complex, higher dimensional systems, the extra representational capacity of the GAN may provide more benefit than can be realized in L96. The computational simplicity of L96 also allows for the production of extremely large training data sets with little compute resources. Higher complexity Earth system model output can provide training set coverage spatially but will be limited temporally by the amount of computational resources available.


\section{Conclusions}
In this study, we have developed an explicitly stochastic GAN framework for the parameterization of sub-grid processes in the Lorenz '96 dynamical system. After testing a wide range of GANs with different noise characteristics, we identified a subset of models that outperform a benchmark bespoke cubic polynomial parameterization.  Returning to the questions posed in the Introduction, we found that this subset of GANs approximates well the joint distribution of the resolved state and the sub-grid-scale tendency.  This model subset also produces the most accurate weather forecasts (i.e. lowest RMSE in the ensemble mean). Some GANs produce reliable forecasts, in which the ensemble spread is a good indicator of the error in the ensemble mean. However, these are not necessarily the GANs that produce the most accurate forecasts. The subset of models with the most accurate weather forecasts produce the most accurate climate simulations, as characterized by probability distributions, space and time correlations, and regime dynamics. However, we note that the GANs which produce skilful weather and climate forecasts were different to those which performed well in ``offline'' mode.

Although the GANs required an iterative development process to maximize model performance and were very sensitive to the noise magnitude and other hyperparameter settings, they do show promise as an approach for stochastic parameterization of physical processes in more complex weather and climate models. Applying other recently developed GAN losses and regularizers \cite{Kurach2019} could help reduce the chance for the GAN to experience a failure mode.

The experiments presented here demonstrate that GANs are a promising machine learning approach for developing stochastic parameterization in complex GCMs. Key lessons learned and unanswered questions include:
\begin{itemize}
    \item While including the tendency from the previous timestep provides a natural approach for building temporal dependence into the parameterization, it can lead to accumulation of error in the forecast, such that local-in-time parameterization should also be considered.
    \item Autocorrelated noise is important for a skillful weather forecast, but appears less important for capturing the climatological distribution.
    \item It is possible that spatial correlations are also important in a higher complexity Earth system model, which could not be assessed here due to the simplicity of the Lorenz `96 system.
    \item It is possible that the noise characteristics could also be learned by the GAN framework to automate the tuning of the stochasticity.
\end{itemize}
Future work will use these lessons to develop machine-learned stochastic parameterization schemes for use in higher complexity Earth system models. GANs of a similar level of complexity to those used for L96 could emulate local effects, such as some warm rain formation processes. Other generative neural network frameworks, such as variational autoencoders, should also be investigated to determine if they can provide similar or better performance with a less sensitive training process.

\acknowledgments
This research started in a working group supported by the Statistical and Applied Mathematical Sciences Institute (SAMSI).  
DJG and HMC were funded by National Center for Atmospheric Research Advanced Study Program Postdoctoral Fellowships and by the National Science Foundation through NCAR's Cooperative Agreement AGS-1852977.
HMC was funded by Natural Environment Research Council grant number NE/P018238/1. AHM acknowledges support from the Natural Sciences and Engineering Research Council of Canada (NSERC) and thanks SAMSI for hosting him in the autumn of 2017. The NCAR Cheyenne supercomputer was used to generate the model runs analyzed in this paper. Sue Ellen Haupt and Joseph Tribbia provided helpful reviews of the paper prior to submission. The source code for this paper can be viewed and downloaded at \url{https://github.com/djgagne/lorenz_gan/}.

\end{document}